# Classification of rotational zero modes in 2D micropolar solids

Dingxin Sun[1], Yi Chen[2, 3*] and Gengkai Hu[1, 3*]

[1]School of Aerospace Engineering, Beijing Institute of Technology (BIT), Beijing 100081, China.

[2]Institute of Applied Physics, Karlsruhe Institute of Technology (KIT), Karlsruhe 76128, Germany.

[3]Marine Science and Technology Domain, Beijing Institute of Technology (BIT), Zhuhai 519088, China.

*Corresponding authors: yi.chen@partner.kit.edu (Y.C.); hugeng@bit.edu.cn (G.H.)

**Abstract**:

Zero modes, which are deformations that cost zero energy, underlie many exotic behaviors in elastic metamaterials. While classical linear Cauchy elasticity explains many of these modes, those linked to the rotations of metamaterial inner components often lie beyond its scope. Micropolar elasticity, which incorporates translation and rotation degrees of freedom, provides a framework for capturing these rotational modes. Herein, we present the first complete symmetry-based classification of zero modes in two-dimensional micropolar solids, with an emphasis on rotation-related modes. Guided by this classification, we construct threefold rotationally symmetric micropolar metamaterials and realize typical rotational micropolar zero modes. We further show that these metamaterials exhibit wave phenomena forbidden in Cauchy continua, including the emergence of three bulk waves in the long-wavelength limit and associated triple refraction, chiral acoustic modes, as well as strong wave anisotropy. All intriguing properties are quantitatively captured by micropolar continuum descriptions, whereas the classical Cauchy continuum theory fails to predict these behaviors, even at a qualitative level. Our results establish a general framework for engineering rotation-based zero modes, opening avenues for designing metamaterials with novel wave properties.

## Introduction

Among the diverse physical mechanisms for elastic, optical, and thermal metamaterials [1-3], zero modes are unique to elastic metamaterials. Physically, zero modes refer to deformations that require zero energy [4-6]. In discrete systems consisting of masses coupled by springs, zero modes generally arise when the number of springs is insufficient to constrain deformations of the systems [5]. Well-known examples are the two-dimensional (2D) Kagome and honeycomb mass-and-spring systems, where each mass, with two translational degrees of freedom, has no more than two bonds on average [7-9]. Mathematically, the Maxwell-Calladine counting rule [10] determines the number of zero modes in such systems. In real metamaterials, ideal springs are replaced by slender beams or weak joints [11-13]. Therefore, zero modes are often realized approximately, yet they still require far less energy than other modes. Zero modes can lead to interesting static and dynamic properties, such as shape-morphing with minimal energy input [14, 15], polarized floppy modes localized at specific edges [16-19] and anomalous cone or Weyl points with robust wave transportation [20, 21].

Many zero modes in metamaterials stem from translations of their inner parts. These can be effectively captured by the classical linear Cauchy elasticity. In the long-wavelength or effective-medium limit, elastic properties are encoded in the Cauchy elasticity matrix. Specifically, each zero eigenvalue of the matrix leads to a zero mode, with the eigenvector describing the corresponding deformation [22]. These deformations are affine deformations with uniform strain and are the so-called Guest-Hutchison modes [23]. Non-zero eigenvalues correspond to hard modes that lead to non-zero strain energy. Following the convention by Milton and Cherkaev [22], three-dimensional (3D) Cauchy materials can be classified, according to the number $N$ of zero eigenvalues, as pentamode ($N = 5$), tetramode ($N = 4$), trimode ($N = 3$), dimode ($N = 2$), monomode ($N = 1$), and nullmode ($N = 0$) materials [24]. In 2D space, we have only dimode, monomode and nullmode materials as the elasticity matrix is $3 \times 3$. These Cauchy elastic materials with zero modes are also named as extremal materials [25-27]. Their bulk wave behaviors [28-30], surface wave properties [31], and polarization behaviors [24, 32] become drastically different from usual solids. A canonical example is the pentamode materials or metamaterials [33-35]. Their five zero modes are all shear deformations, and the only hard mode is hydrostatic compression or expansion [29, 36]. Therefore, pentamode metamaterials transmit exclusively longitudinal waves as fluids [37-39], but can be engineered to exhibit extremely wave anisotropy [40, 41], which are especially useful in realizing broadband underwater acoustic cloaks [42, 43, 28].

Incorporating rotations into metamaterials substantially enriches their possible deformations and properties [44, 45], such as vanishing bulk modulus in dilational metamaterials induced by rotation-dilation coupling [46] or push-to-twist effects [45] and acoustical activity [47-49] in chiral metamaterials that lack centrosymmetry. Leveraging rotation-based zero modes expands the possibilities further. One recent example is to achieve the highly unusual roton-like dispersion [50, 51], which is proposed to understand the exotic superfluid Helium-4 by physicist Landau [52]. This dispersion is featured with a local minimum of the wave energy versus the wavenumber [53-55] and has been realized in various correlated quantum systems at cryogenic temperatures [56, 57]. Rotational zero modes [58], among others, including nonlocality [54, 59], offer an opportunity to study roton-like dispersions at room temperature. Such highly unusual dispersion relations can lead to broadband negative refraction, multiple refraction, zero group velocity and so on [53, 55]. In the static regime, the intriguing dispersion further leads to a violation of the Saint-Venant's principle, i.e., mechanical responses become extremely sensitive to boundary conditions even at far positions [60, 61]. Additionally, rotational zero modes provide a mechanism to design metamaterials with asymmetric effective elasticity tensors [62-64], where asymmetric shear stress is balanced by rotation inertia [65, 63]. Following asymmetric transformation elasticity [66], these metamaterials have been adopted to design elastodynamic cloaks with broadband effectiveness [67-69].

For metamaterials with rotational zero modes as above, the Cauchy effective-medium often cannot correctly capture their anomalous behaviors. Micropolar elasticity, also known as Cosserat theory, is a suitable framework to model rotation-related motions [45, 70, 71]. This theory is a generalization of Cauchy elasticity by assigning independent micro-rotations to each material point [72]. Three fourth-order micropolar elasticity tensors are introduced in the theory, in contrast to only one fourth-order tensor in Cauchy elasticity. For micropolar solids without any spatial symmetry, 171 independent material parameters are needed [72]. Even for isotropic chiral micropolar solids, 9 instead of 2 material parameters are required. Micropolar elasticity has long been used to model size effects and nonlocal responses in complex materials, such as bones or foams with porous features [70]. Recently, this theory has been extensively studied in the context of metamaterials, particularly for those with pronounced inner-rotations [73-76]. For instance, the above-mentioned chiral effects, push-to-twist coupling and acoustical activity, in 3D chiral metamaterials are perfectly modeled by using micropolar effective-medium parameters can be obtained using different approaches, including by fitting metamaterials dispersion rela [76-78]. By extending the theory to cover extremal micropolar solids, the roton-like dispersion relation and asymmetric elasticity mentioned above can also be explained [79]. Apart from a few examples, extremal micropolar solids and micropolar zero modes have remained largely unexplored.

In this work, we present the first comprehensive classification of zero modes in micropolar solids, with an emphasis on those linked to rotations. In micropolar elasticity, deformations are measured by both micropolar strains, related to gradient of displacements and micro-rotation, and micropolar curvature, related to gradient of micro-rotations. Here, we focus on 2D micropolar elasticity. The classification is already demanding as deformations lie in a space with dimension 6, the same number as for 3D Cauchy elasticity. For 3D micropolar elasticity, the dimension of the deformation space becomes 18. We first introduce a set of 6 deformation bases, which provides a clear physical picture for interpreting general deformations [80]. The six deformation bases themselves or coupled bases could be zero modes, depending on the material symmetry group. We remark that previous study have investigated the micropolar elasticity matrices of materials with different symmetries [81-83]. These work have mainly focused on revealing possible coupling between different deformations, e.g., coupling between twisting and stretching. Our work instead has identified allowed zero modes within micropolar materials in different symmetry groups. Additionally, our chosen deformation bases are slightly different from previous work, leading to a simpler structure of the transformed elasticity matrix as shown later. This simpler transformed elasticity matrix makes it easier to classify zero modes as decoupled or coupled zero modes. We classify zero modes in micropolar solids according to the eight crystal symmetry groups $\mathrm{C_n}$ and $\mathrm{C_{nv}}$ $(\mathrm{n} = 6, 4, 3, 2)$. The symmetry notation $\mathrm{C_n}$ denotes $\mathrm{n}$-fold rotational symmetry, i.e., a material is invariant under a rotation of $2\pi/n$. Materials belonging to the $\mathrm{C_{nv}}$ symmetry additionally possess mirror-reflection symmetry. We further propose discrete micropolar metamaterials with $\mathrm{C_3}$ and $\mathrm{C_{3v}}$ symmetry to construct several typical rotation-based zero modes. By engineering rotational zero modes, we show that three acoustic modes and complicated mode hybridization can occur even in the low-frequency limit. According to Cauchy-continuum theory, these high-symmetric metamaterials are isotropic in the long-wavelength limit. In sharp contrast, we demonstrate strongly anisotropic wave propagation, which agrees with micropolar-continuum prediction. Our results thus also demonstrate the efficiency of micropolar elasticity in modeling rotation-based zero modes.

The remainder of this paper is organized as follows. In Section 2, we introduce the basics of 2D micropolar elasticity and define the deformation bases. Then we present the classification of zero modes in micropolar solids, organized according to crystal symmetry groups, including achiral groups, $\mathrm{C_{6v}}$, $\mathrm{C_{4v}}$, $\mathrm{C_{3v}}$, $\mathrm{C_{2v}}$ and chiral ones, $\mathrm{C_6}$, $\mathrm{C_4}$, $\mathrm{C_3}$, $\mathrm{C_2}$. Allowed zero modes are identified. In Section 3, we design discrete micropolar metamaterials in $\mathrm{C_{3v}}$ and $\mathrm{C_3}$ groups to construct symmetry-allowed zero modes tied to micro-rotations. Micropolar effective-medium parameters are analytically derived for the metamaterials. We obtain excellent agreement between results from the micropolar-continuum model and the direct metamaterial calculations, including dispersion relations, wave polarization, bulk wave propagations, and wave refractions at interfaces. We also show that the Cauchy-continuum model can hardly reproduce the metamaterial results. Section 4 concludes the study.

## Classification of zero modes in micropolar solids

In this section, we provide a brief introduction to micropolar-continuum theory and introduce a set of deformation bases for characterizing zero modes. Here, we consider a Cartesian coordinate system, with the coordinate variables $x_\alpha$, $\alpha = 1, 2$. The two orthonormal bases are denoted as $\mathbf{e}_\alpha$. In 2D micropolar elasticity, each material point has two translation degrees of freedom (DOFs), $u_\alpha$, and one micro-rotation DOF $\phi$. It should be noted that the micro-rotation $\phi$ is independent from the macro-rotation, i.e., the anti-symmetric part of the displacement gradient $1/2(\partial u_2/\partial x_1 - \partial u_1/\partial x_2)$. The deformation is characterized by the micropolar strain tensor $\varepsilon_{\alpha\beta}$ and the micropolar curvature $\kappa_\alpha$

$$\varepsilon_{\alpha\beta} = \frac{\partial u_\beta}{\partial x_\alpha} - \epsilon_{\alpha\beta}\phi, \quad \kappa_\alpha = \frac{\partial \phi}{\partial x_\alpha}, \tag{1}$$

where $\epsilon_{\alpha\beta}$ represents the Levi-Civita tensor. All Greek indices range from 1 to 2. Note that the micro-rotation is involved in the micropolar strain. Since a rigid rotation leads to equal macro-rotation and micro-rotation, the micropolar strain defined in Eq. (1) ensures the strain is zero under rigid rotation.

Owing to the intrinsic micro-rotation DOF, not only stress but also couple stress, i.e., moment of force, are introduced. We denote the micropolar stress tensor and the micropolar couple stress as $\sigma_{\alpha\beta}$ and $m_\alpha$, respectively. Here, $\sigma_{\alpha\beta}$ represents surface force along $\mathbf{e}_\beta$ on a unit area with surface norm $\mathbf{e}_\alpha$, and $m_\alpha$ stands for moment of force on a unit area with surface norm $\mathbf{e}_\alpha$. The balance law for linear momentum and angular momentum in the frequency domain reads

$$-\omega^2 \rho u_\alpha = \frac{\partial \sigma_{\beta\alpha}}{\partial x_\beta}, \quad -\omega^2 I \phi = \frac{\partial m_\alpha}{\partial x_\alpha} + \epsilon_{\alpha\beta}\sigma_{\alpha\beta}. \tag{2}$$

In which, $\rho$ is the mass density, $I$ represents the micro-rotation inertia density (micro-rotation inertia per unit volume), and $\omega$ is the angular frequency. In Cauchy theory, the stress tensor is symmetric due to the balance of angular momentum. Here, the micropolar stress tensor is generally not symmetric. Its asymmetric part is balanced by the couple stress and the micro-rotation inertia. This character enables designing elastodynamic cloaks based on the asymmetric transformation method [65].

In linear micropolar elasticity, we have a quadratic strain energy density function in terms of the micropolar strain and the micropolar curvature

$$w = \frac{1}{2}\varepsilon_{\alpha\beta} C_{\alpha\beta\gamma\delta} \varepsilon_{\gamma\delta} + \varepsilon_{\alpha\beta} B_{\alpha\beta\gamma} \kappa_\gamma + \frac{1}{2}\kappa_\alpha A_{\alpha\beta} \kappa_\beta. \tag{3}$$

The micropolar stress and the couple stress are derived by differentiating the strain energy density function $w$ with respect to the strain and the curvature, respectively,

$$\sigma_{\alpha\beta} = C_{\alpha\beta\gamma\delta}\varepsilon_{\gamma\delta} + B_{\alpha\beta\gamma}\kappa_\gamma, \quad m_\alpha = A_{\alpha\beta}\kappa_\beta + B_{\beta\gamma\alpha}\varepsilon_{\beta\gamma}. \tag{4}$$

The fourth-order micropolar elasticity tensor $C_{\alpha\beta\gamma\delta}$ exhibits major symmetry, $C_{\alpha\beta\gamma\delta} = C_{\gamma\delta\alpha\beta}$, but has no minor symmetry. Likewise, the second-order tensor $A_{\alpha\beta}$ is symmetric, i.e., $A_{\alpha\beta} = A_{\beta\alpha}$. The third-order tensor $B_{\alpha\beta\gamma}$ couples micropolar curvature to stress and couple stress to strain, and is also called the coupling tensor [72].

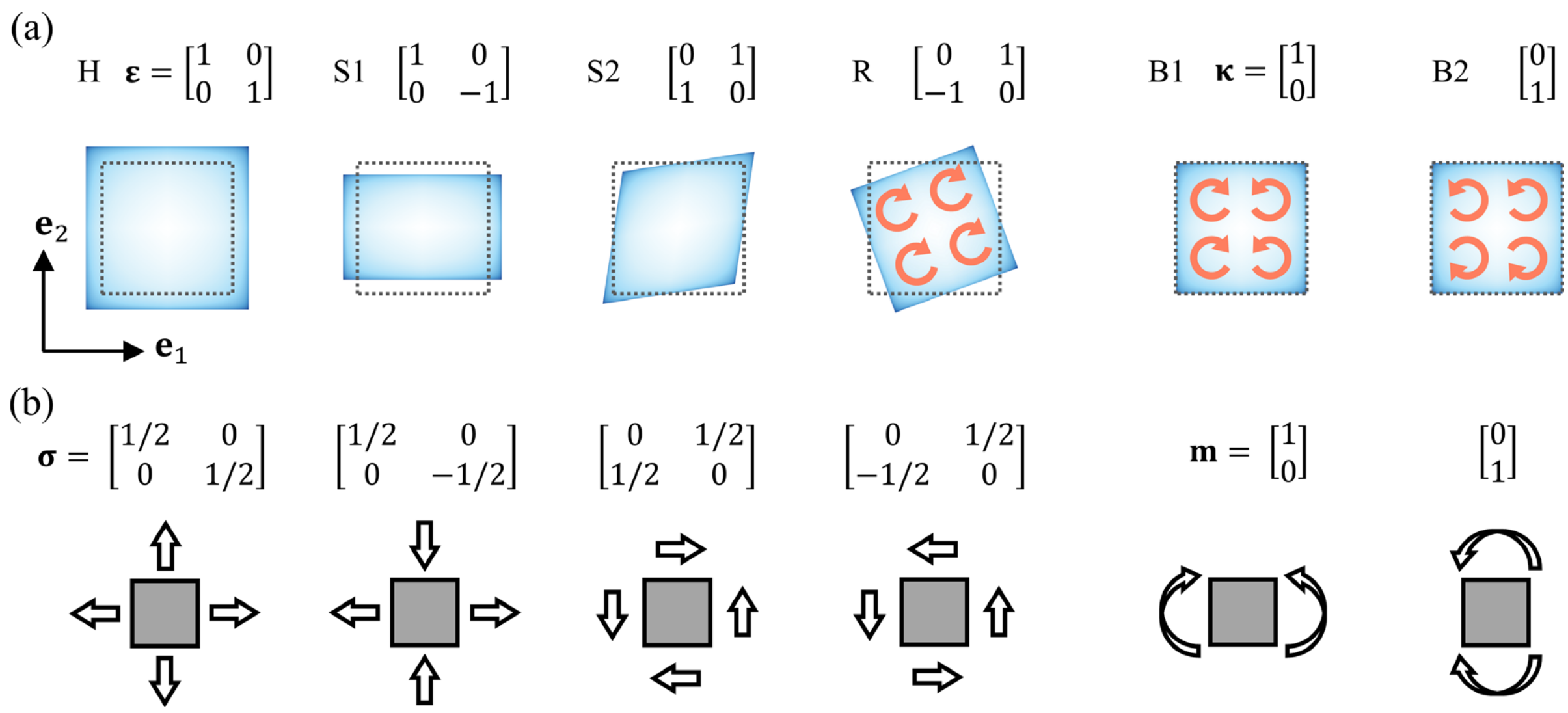

**Figure 1**. **Bases for micropolar strain, curvature, stress, and couple stress**. (a) Illustration of six deformation bases. The first four correspond to non-zero micropolar strain, $\boldsymbol{\varepsilon}^{\mathrm{o}}, \boldsymbol{\varepsilon}^{\mathrm{d}}, \boldsymbol{\varepsilon}^{\mathrm{s}}$ and $\boldsymbol{\varepsilon}^{\mathrm{a}}$, as indicated by the matrices. The dashed lines represent the edges of an undeformed infinitesimal square element. The blue region stands for the deformed shape of the infinitesimal square element. Physically, the four deformations represent hydrostatic expansion, shear deformation, pure shear,

and pure rotation, respectively. The four deformation bases are denoted by letters "H", "S1", "S2", and "R", respectively. The last two denote deformations resulting from non-zero micropolar curvature, as illustrated by the corresponding vectors. They are denoted by letters "B1" and "B2". Micro-rotation is represented by arrows. (b) Illustration of four micropolar stress and two couple stress bases. Arrows indicate traction and curved arrows represent the moment of force.

For a general micropolar strain tensor $\boldsymbol{\varepsilon}$, we can decompose it by using the following four bases

$$\boldsymbol{\varepsilon} = \nabla\mathbf{u} - \phi\boldsymbol{\varepsilon}^{\mathrm{a}} = \varepsilon_{\mathrm{o}}\boldsymbol{\varepsilon}^{\mathrm{o}} + \varepsilon_{\mathrm{d}}\boldsymbol{\varepsilon}^{\mathrm{d}} + \varepsilon_{\mathrm{s}}\boldsymbol{\varepsilon}^{\mathrm{s}} + \varepsilon_{\mathrm{a}}\boldsymbol{\varepsilon}^{\mathrm{a}}, \tag{5}$$

in which,

$$\begin{cases} \boldsymbol{\varepsilon}^{\mathrm{o}} = \mathbf{e}_1 \otimes \mathbf{e}_1 + \mathbf{e}_2 \otimes \mathbf{e}_2, & \varepsilon_{\mathrm{o}} = \dfrac{u_{1,1} + u_{2,2}}{2}, \\ \boldsymbol{\varepsilon}^{\mathrm{d}} = \mathbf{e}_1 \otimes \mathbf{e}_1 - \mathbf{e}_2 \otimes \mathbf{e}_2, & \varepsilon_{\mathrm{d}} = \dfrac{u_{1,1} - u_{2,2}}{2}, \\ \boldsymbol{\varepsilon}^{\mathrm{s}} = \mathbf{e}_1 \otimes \mathbf{e}_2 + \mathbf{e}_2 \otimes \mathbf{e}_1, & \varepsilon_{\mathrm{s}} = \dfrac{u_{1,2} + u_{2,1}}{2}, \\ \boldsymbol{\varepsilon}^{\mathrm{a}} = \mathbf{e}_1 \otimes \mathbf{e}_2 - \mathbf{e}_2 \otimes \mathbf{e}_1, & \varepsilon_{\mathrm{a}} = \dfrac{u_{2,1} - u_{1,2}}{2} - \phi, \end{cases} \tag{6}$$

The four deformation bases are mutually orthogonal in the sense that $\boldsymbol{\varepsilon}^i : \boldsymbol{\varepsilon}^j = 0$ for $i \neq j \in \{\mathrm{o}, \mathrm{d}, \mathrm{s}, \mathrm{a}\}$. In Fig. 1(a), we illustrate the four micropolar strain bases, i.e., $\boldsymbol{\varepsilon}^{\mathrm{o}}$, $\boldsymbol{\varepsilon}^{\mathrm{d}}$, $\boldsymbol{\varepsilon}^{\mathrm{s}}$ and $\boldsymbol{\varepsilon}^{\mathrm{a}}$. Physically, the four deformations correspond to hydrostatic expansion, shear deformation in a rotated frame, pure shear deformation, and pure rotation, respectively. For convenience, we denote these deformation bases by the symbols "H", "S1", "S2", and "R", respectively, throughout the main text. A general micropolar curvature can be decomposed according to two orthogonal basis vectors

$$\boldsymbol{\kappa} = \kappa_1\boldsymbol{\kappa}^1 + \kappa_2\boldsymbol{\kappa}^2, \qquad \boldsymbol{\kappa}^1 = (1{,}0), \qquad \boldsymbol{\kappa}^2 = (0{,}1). \tag{7}$$

We denote the basic deformations for the two vectors by "B1" and "B2", respectively. It is worth noting that micro-rotation cannot be represented by a change in the shape of the infinitesimal element and is instead depicted by arrows (Fig. 1(a)).

Likewise, we can decompose a micropolar stress tensor into following four stress bases

$$\boldsymbol{\sigma} = \sigma_{\mathrm{o}}\boldsymbol{\sigma}^{\mathrm{o}} + \sigma_{\mathrm{d}}\boldsymbol{\sigma}^{\mathrm{d}} + \sigma_{\mathrm{s}}\boldsymbol{\sigma}^{\mathrm{s}} + \sigma_{\mathrm{a}}\boldsymbol{\sigma}^{\mathrm{a}}, \tag{8}$$

in which,

$$\begin{cases} \boldsymbol{\sigma}^{\mathrm{o}} = \dfrac{\mathbf{e}_1 \otimes \mathbf{e}_1 + \mathbf{e}_2 \otimes \mathbf{e}_2}{2}, & \sigma_{\mathrm{o}} = \sigma_{11} + \sigma_{22}, \\ \boldsymbol{\sigma}^{\mathrm{d}} = \dfrac{\mathbf{e}_1 \otimes \mathbf{e}_1 - \mathbf{e}_2 \otimes \mathbf{e}_2}{2}, & \sigma_{\mathrm{d}} = \sigma_{11} - \sigma_{22}, \\ \boldsymbol{\sigma}^{\mathrm{s}} = \dfrac{\mathbf{e}_1 \otimes \mathbf{e}_2 + \mathbf{e}_2 \otimes \mathbf{e}_1}{2}, & \sigma_{\mathrm{s}} = \sigma_{12} + \sigma_{21}, \\ \boldsymbol{\sigma}^{\mathrm{a}} = \dfrac{\mathbf{e}_1 \otimes \mathbf{e}_2 - \mathbf{e}_2 \otimes \mathbf{e}_1}{2}, & \sigma_{\mathrm{a}} = \sigma_{12} - \sigma_{21}. \end{cases} \tag{9}$$

Two bases for the micropolar couple stress are defined as $\mathbf{m}^1 = (1, 0)$ and $\mathbf{m}^2 = (0{,}1)$. The physical meaning of these stress and couple stress bases is illustrated in Fig. 1(b).

The micropolar constitutive law in Eq. (4) can be rewritten in terms of the bases defined above,

$$\begin{pmatrix} \sigma_\mathrm{o} \\ \sigma_\mathrm{d} \\ \sigma_\mathrm{s} \\ \sigma_\mathrm{a} \\ m_1 \\ m_2 \end{pmatrix} = \begin{pmatrix} \hat{C}_{11} & \hat{C}_{12} & \hat{C}_{13} & \hat{C}_{14} & \hat{B}_{11} & \hat{B}_{12} \\ & \hat{C}_{22} & \hat{C}_{23} & \hat{C}_{24} & \hat{B}_{21} & \hat{B}_{22} \\ & & \hat{C}_{33} & \hat{C}_{34} & \hat{B}_{31} & \hat{B}_{32} \\ & & & \hat{C}_{44} & \hat{B}_{41} & \hat{B}_{42} \\ & \mathrm{sym} & & & \hat{A}_{11} & \hat{A}_{12} \\ & & & & & \hat{A}_{22} \end{pmatrix} \begin{pmatrix} \varepsilon_\mathrm{o} \\ \varepsilon_\mathrm{d} \\ \varepsilon_\mathrm{s} \\ \varepsilon_\mathrm{a} \\ \kappa_1 \\ \kappa_2 \end{pmatrix}. \tag{10}$$

The transformed elasticity matrix in Eq. (10) has 21 independent constants. These constants are determined by the micropolar tensors in Eq. (4) and their expressions are given in **Appendix A**.

**Table 1. Classification of micropolar zero modes according to material symmetries**. The transformed elasticity matrix is listed in the second column. Symbol "-" means a zero mode that couples multiple basic deformations, whereas "&" means several zero modes occurs simultaneously due to symmetry. The designed metamaterials in next section support zero modes marked by "*".

| Symmetry | Transformed Elasticity Matrix | Symmetry-Allowed Zero Modes | |
|---|---|---|---|
| $C_{6v}$ | $\begin{pmatrix} \hat{C}_{11} & 0 & 0 & 0 & 0 & 0 \\ & \hat{C}_{22} & 0 & 0 & 0 & 0 \\ & & \hat{C}_{22} & 0 & 0 & 0 \\ & & & \hat{C}_{44} & 0 & 0 \\ & \mathrm{sym} & & & \hat{A}_{11} & 0 \\ & & & & & \hat{A}_{11} \end{pmatrix}$ | Decoupled | H, S1 & S2, R, B1 & B2 |
| | | Coupled | None |
| $C_6$ | $\begin{pmatrix} \hat{C}_{11} & 0 & 0 & \hat{C}_{14} & 0 & 0 \\ & \hat{C}_{22} & 0 & 0 & 0 & 0 \\ & & \hat{C}_{22} & 0 & 0 & 0 \\ & & & \hat{C}_{44} & 0 & 0 \\ & \mathrm{sym} & & & \hat{A}_{11} & 0 \\ & & & & & \hat{A}_{11} \end{pmatrix}$ | Decoupled | H, S1 & S2, R, B1 & B2 |
| | | Coupled | H-R: $\hat{C}_{11}\hat{C}_{44} = \hat{C}_{14}^2$ |
| $C_{3v}$ | $\begin{pmatrix} \hat{C}_{11} & 0 & 0 & 0 & 0 & 0 \\ & \hat{C}_{22} & 0 & 0 & \hat{B}_{21} & \hat{B}_{22} \\ & & \hat{C}_{22} & 0 & \hat{B}_{22} & -\hat{B}_{21} \\ & & & \hat{C}_{44} & 0 & 0 \\ & \mathrm{sym} & & & \hat{A}_{11} & 0 \\ & & & & & \hat{A}_{11} \end{pmatrix}$ | Decoupled | H, S1&S2, R*, B1 & B2 |
| | | Coupled | S1-S2-B1 & S1-S2-B2*: $\hat{C}_{22}\hat{A}_{11} = \hat{B}_{21}^2 + \hat{B}_{22}^2$ |
| $C_3$ | $\begin{pmatrix} \hat{C}_{11} & 0 & 0 & \hat{C}_{14} & 0 & 0 \\ & \hat{C}_{22} & 0 & 0 & \hat{B}_{21} & \hat{B}_{22} \\ & & \hat{C}_{22} & 0 & \hat{B}_{22} & -\hat{B}_{21} \\ & & & \hat{C}_{44} & 0 & 0 \\ & \mathrm{sym} & & & \hat{A}_{11} & 0 \\ & & & & & \hat{A}_{11} \end{pmatrix}$ | Decoupled | H, S1&S2, R, B1 & B2 |
| | | Coupled | H-R*, S1-S2-B1 & S1-S2-B2 |

In Tables 1 and 2, we classify allowed micropolar zero modes according to material symmetry groups. The second column shows the transformed micropolar elasticity matrix. The third column lists symmetry-allowed zero modes. A *decoupled* zero mode refers to a mode associated with a single deformation basis defined in Eq. (6), whereas a *coupled* zero mode corresponds to a linear combination of multiple deformation bases. In general, a zero mode arises when a principal minor of the transformed elasticity matrix has a vanishing determinant [22]. As an example, the transformed elasticity matrix of a material with $C_{6v}$ symmetry is diagonal. In this case, the hydrodynamic expansion ("H") becomes a zero mode if $\hat{C}_{11} = 0$. For a material with $C_6$ symmetry, the condition $\hat{C}_{11}\hat{C}_{44} - \hat{C}_{14}^2 = 0$ gives rise to a coupled zero mode that is a linear combination of the hydrodynamic expansion ("H") and the pure rotation ("R").

In this work, we focus on zero modes associated with the micro-rotation DOF. As shown in Table 1, materials with $C_3$ or $C_{3v}$ symmetry is capable to realize multiple typical rotational zero modes. For example, with $C_{3v}$-symmetric materials, we can achieve either a pure rotational zero mode ("R") or a coupled zero mode involving rotation gradients and shear deformations ("S1–S2–B1"). By reducing the symmetry to $C_3$, a coupled zero mode combining the hydrodynamic expansion and the pure rotation ("H–R") is possible. In the following metamaterial designs, we therefore restrict our attention to these two symmetry classes.

**Table 2. Same as Table 1 but for remaining symmetric groups $C_{4v}$, $C_4$, $C_{2v}$ and $C_2$.**

| Symmetry | Transformed Elasticity Matrix | Symmetry-Allowed Zero Modes | |
|---|---|---|---|
| $C_{4v}$ | $\begin{pmatrix} \hat{C}_{11} & 0 & 0 & 0 & 0 & 0 \\ & \hat{C}_{22} & 0 & 0 & 0 & 0 \\ & & \hat{C}_{33} & 0 & 0 & 0 \\ & & & \hat{C}_{44} & 0 & 0 \\ & \text{sym} & & & \hat{A}_{11} & 0 \\ & & & & & \hat{A}_{11} \end{pmatrix}$ | Decoupled | H, S1, S2, R, B1 & B2 |
| | | Coupled | None |
| $C_4$ | $\begin{pmatrix} \hat{C}_{11} & 0 & 0 & \hat{C}_{14} & 0 & 0 \\ & \hat{C}_{22} & \hat{C}_{23} & 0 & 0 & 0 \\ & & \hat{C}_{33} & 0 & 0 & 0 \\ & & & \hat{C}_{44} & 0 & 0 \\ & \text{sym} & & & \hat{A}_{11} & 0 \\ & & & & & \hat{A}_{11} \end{pmatrix}$ | Decoupled | H, S1, S2, R, B1 & B2 |
| | | Coupled | H-R, S1-S2: $\hat{C}_{22}\hat{C}_{33} = \hat{C}_{23}^2$ |
| $C_{2v}$ | $\begin{pmatrix} \hat{C}_{11} & \hat{C}_{12} & 0 & 0 & 0 & 0 \\ & \hat{C}_{22} & 0 & 0 & 0 & 0 \\ & & \hat{C}_{33} & \hat{C}_{34} & 0 & 0 \\ & & & \hat{C}_{44} & 0 & 0 \\ & \text{sym} & & & \hat{A}_{11} & \hat{A}_{12} \\ & & & & & \hat{A}_{22} \end{pmatrix}$ | Decoupled | H, S1, S2, R, B1, B2 |
| | | Coupled | H-S1: $\hat{C}_{11}\hat{C}_{22} = \hat{C}_{12}^2$,<br>S2-R: $\hat{C}_{33}\hat{C}_{44} = \hat{C}_{34}^2$,<br>B1-B2: $\hat{A}_{11}\hat{A}_{22} = \hat{A}_{12}^2$ |
| $C_2$ | $\begin{pmatrix} \hat{C}_{11} & \hat{C}_{12} & \hat{C}_{13} & \hat{C}_{14} & 0 & 0 \\ & \hat{C}_{22} & \hat{C}_{23} & \hat{C}_{24} & 0 & 0 \\ & & \hat{C}_{33} & \hat{C}_{34} & 0 & 0 \\ & & & \hat{C}_{44} & 0 & 0 \\ & \text{sym} & & & \hat{A}_{11} & \hat{A}_{12} \\ & & & & & \hat{A}_{22} \end{pmatrix}$ | Decoupled | H, S1, S2, R, B1, B2 |
| | | Coupled | H-R, H-S, S1-S2,<br>S1-R, S2-R, B1-B2 |

### Metamaterial design and numerical simulations

The twisted (regular) Kagome lattice is a typical lattice belonging to the $C_3$ ($C_{3v}$) symmetry class. Based on this lattice, we introduce several modifications to obtain the discrete metamaterial model shown in Fig. 2(a). Specifically, half of the triangles in the Kagome lattice are filled with solid mass blocks (gray triangles), whose two translation DOFs and one rotation DOF are used to mimic the three DOFs in micropolar elasticity. The other half of the triangles are implemented using springs (black lines). In addition, we introduce an extra set of springs (blue lines) to tune the effective coupling elasticity matrix $\widehat{\mathbf{B}}$. A unit cell is highlighted in light orange, and the two lattice vectors are given by $\mathbf{a}_1 = (a, 0)$ and $\mathbf{a}_2 = (a/2,\ \sqrt{3}a/2)$. This metamaterial supports different zero modes depending on the geometric parameters and the spring constants. In the following, we briefly introduce the main steps to derive the dynamic equations and focus more on the analysis of the zero modes and wave properties. Our analysis is limited to the linear elasticity regime or small deformations.

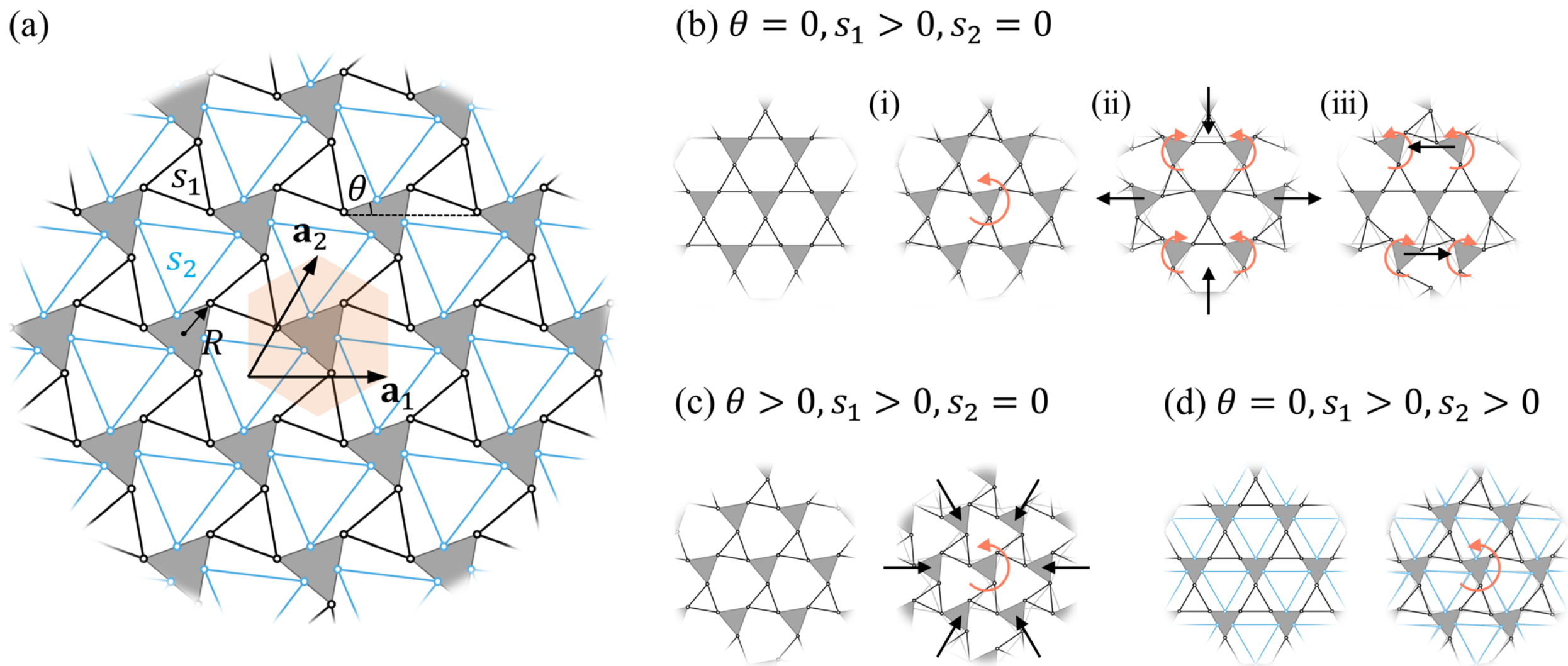

**Figure 2**. **A unified metamaterial design exhibiting multiple rotation-related zero modes**. (a) A discrete triangular lattice metamaterial, with a unit cell highlighted in orange, composed of mass blocks (gray triangles) connected by linear axial springs (black and blue lines). Each mass block has the same mass $M$ and moment of inertia $J = MR^2/4$. Geometry parameters are indicated and $R = a/(2\sqrt{3}\cos\theta)$, with $a$ being the lattice constant. The two indicated lattice vectors are $\mathbf{a}_1 = (a, 0)$ and $\mathbf{a}_2 = (a/2,\ \sqrt{3}a/2)$. Different choices of the twist angle $\theta$ and the spring constant $s_2$ and $s_1$ lead to different zero modes. (b) A metamaterial with three zero modes: (i) a pure rotation zero mode, and (ii), (iii) two coupled zero modes containing rotation gradient and shear deformation. (c) A metamaterial with a coupled zero mode consisting of hydrostatic expansion and pure rotation. (d) A metamaterial with only a pure rotational zero mode.

For the periodic metamaterial, we can label its unit cell by two integers $(p,\ q)$ after choosing a reference unit cell $(0, 0)$. The unit cell $(p,\ q)$ locates at the position $p\mathbf{a}_1 + q\mathbf{a}_2$ relatively to the reference one $(0, 0)$. We denote the center displacement of the mass block in unit cell $(p,\ q)$ and its rotation as $\mathbf{u}_{p,q} = (u_{p,q},\ v_{p,q})$ and $\phi_{p,q}$, respectively. The Lagrangian related to the DOFs for the unit cell $(p,\ q)$ can be written as

$$L_{p,q} = T_{p,q} - V_{p,q}, \tag{11}$$

with $T_{p,q}$ representing the kinetic energy of the mass block in the unit cell $(p,\ q)$, and $V_{p,q}$ representing the potential energy in the six black springs and six blue springs connected to the central mass block in the unit cell $(p,\ q)$. The governing equations for the displacement and rotation can be derived as

$$\frac{\mathrm{d}}{\mathrm{d}t}\left(\frac{\partial L_{p,q}}{\partial \dot{u}_{p,q}}\right) - \frac{\partial L_{p,q}}{\partial u_{p,q}} = 0, \qquad \frac{\mathrm{d}}{\mathrm{d}t}\left(\frac{\partial L_{p,q}}{\partial \dot{v}_{p,q}}\right) - \frac{\partial L_{p,q}}{\partial v_{p,q}} = 0, \qquad \frac{\mathrm{d}}{\mathrm{d}t}\left(\frac{\partial L_{p,q}}{\partial \dot{\phi}_{p,q}}\right) - \frac{\partial L_{p,q}}{\partial \phi_{p,q}} = 0. \tag{12}$$

In calculating the dispersion relations of the metamaterials, we assume Bloch wave solution $\mathbf{u}_{p,q} = \hat{\mathbf{u}}_0 \exp(\mathrm{i}\mathbf{k}\cdot\mathbf{r} - \mathrm{i}\omega t) = (\hat{u}_0, \hat{v}_0)\exp(\mathrm{i}\mathbf{k}\cdot(p\mathbf{a}_1 + q\mathbf{a}_2) - \mathrm{i}\omega t)$ , $\phi_{p,q} = \hat{\phi}_0 \exp(i\mathbf{k}\cdot(p\mathbf{a}_1 + q\mathbf{a}_2) - \mathrm{i}\omega t)$, with i representing imaginary unit, $\mathbf{k}$ being the Bloch wave vector. The unknown constants $\hat{u}_0, \hat{v}_0$ and $\hat{\phi}_0$ stand for the displacement and rotation of the Bloch mode. Substituting the Bloch wave form into Eq. (12) leads to an eigenvalue problem. For a given Bloch wavevector $\mathbf{k}$, the eigenfrequency $\omega$ and the eigenmode $(\hat{u}_0, \hat{v}_0, \hat{\phi}_0)$ are solved.

The first case (Fig. 2(b), $\theta = 0$ and $s_2 = 0$) recovers the well-known Kagome lattice with $C_{3v}$ symmetry [5]. We remark that this metamaterial is a Maxwell lattice or isostatic lattice, with an equal number of DOFs and bonds [6, 17, 18]. Previously, a micro-twist continuum theory, with intentionally introduced a rotational DOF, has been proposed to model this metamaterial [8, 9]. In this paper, we will show that the well-established micropolar elasticity can be used to understand and predict its wave behaviors.

From the potential energy (see details in **Appendix B**), it can be verified that this metamaterial has three zero modes. The first zero mode (Fig. 2(b)(i)) is a pure rotation zero mode, with all gray triangles having the same rotation, $\phi_{p,q} = \phi_0$. This zero mode can be intuitively understood since a uniform rotation leads to zero elongation of all springs. The second (Fig. 2(b)(ii)) and third (Fig. 2(b)(iii)) zero modes are both coupled zero modes with shear deformation and gradient rotation. From an micropolar effective-medium description, the three zero modes are represented by $(\varepsilon_\mathrm{o}, \varepsilon_\mathrm{d}, \varepsilon_\mathrm{s}, \varepsilon_\mathrm{a}, \kappa_1, \kappa_2) = (0,0,0,1,0,0)$ , $(0,1,0,0,2/R,0)$ and $(0,0,1,0,0,-2/R)$ , respectively (see details in **Appendix B**).

The twisted Kagome lattice with broken mirror-reflection symmetry is obtained if a non-zero twist angle is chosen [5]. This metamaterial supports a zero mode comprising both rotation and expansion (Fig. 2(c), $\theta = 10^\mathrm{o}$, $s_2 = 0$). Unlike for the metamaterial in Fig. 2(b), the expansion here is necessary to maintain zero stretching of all springs. In micropolar theory, this zero mode is a coupled zero mode represented by $(\varepsilon_\mathrm{o}, \varepsilon_\mathrm{d}, \varepsilon_\mathrm{s}, \varepsilon_\mathrm{a}, \kappa_1, \kappa_2) = (\sin\theta, 0, 0, \cos\theta, 0, 0)$ (see details in **Appendix B**). In the Cauchy-continuum description, this zero mode is reflected by a Poisson's ratio of $-1$. It leads to an equal phase velocity for longitudinal wave and transverse wave, as will be shown later. The coupled zero mode is not only valid for small deformation but exists for general angles $\theta$. This lattice can perfectly fold into a compact state along this deformation trajectory with zero energy cost [17, 18].

For the metamaterial with both types of springs (Fig. 2(d), $\theta = 0$, $s_2 = 40s_1$), it has only one pure rotation zero mode beyond Cauchy-continuum description. In fact, when the metamaterial is homogenized as a Cauchy-continuum medium, its effective elasticity matrix (see Eq. (C9) in **Appendix C**) has no zero mode since any non-zero strain will cost energy. Among all three examples, the current one is the most mechanically stable design. This metamaterial provides a realistic example for practical fabrication and we focus on this one in the wave simulations presented later.

Below, we show that the wave behaviors of the metamaterials above with rotational zero modes can be effectively modeled by using micropolar-continuum theory, while Cauchy-continuum is insufficient to describe various aspects. Micropolar and Cauchy effective-medium parameters can be obtained using different approaches, including by fitting metamaterials dispersion relations to theory [40], by homogenizing the discrete dynamic governing equations [63], or by considering strain-energy equivalence principle [62]. The last one is used for our paper (see details in **Appendix C**) due to its robustness in handling different geometries. We obtain the strain energy density by expanding the displacement and rotation to first order, as shown in Eq. (C2) in **Appendix C**. This expansion provides good approximation in the small-wavenumber regime, as the band structure shows later. Higher-order expansion is required to obtain agreement over a larger wavenumber range; however, the resulting continuum theory goes beyond micropolar elasticity. The effective mass density $\rho$ and micro-rotation inertia density $I$ are obtained by averaging the mass and the moment of inertia over a unit cell, $\rho = M/S_{\text{cell}}$ and $I = J/S_{\text{cell}}$, with $S_{\text{cell}} = \sqrt{3}a^2/2$ being the unit cell area.

Figure 3(a) shows the dispersion relations of the discrete Kagome metamaterial illustrated in Fig. 2(b). The normalization frequency is $\omega_0 = \sqrt{(s_1 + s_2)/M}$. The Bloch wave vector is swept along the high-symmetry directions, as the inset shows. The high symmetric positions are $\mathbf{K} = (4\pi/(3a)$ , 0), and $\mathbf{M} = \left(\pi/a,\ \pi/(\sqrt{3}a)\right)$. For comparison, dispersion relations obtained from the effective micropolar-continuum and the classical Cauchy-continuum are also shown as solid and dashed lines, respectively. False colors are used to encode the polarization of the Bloch modes for each branch. For a Bloch mode of the form $\mathbf{u}_{p,q} = \hat{\mathbf{u}}_0 \exp(\mathrm{i}\mathbf{k} \cdot (p\mathbf{a}_1 + q\mathbf{a}_2) - \mathrm{i}\omega t)$, $\phi_{p,q} = \hat{\phi}_0 \exp(i\mathbf{k} \cdot (p\mathbf{a}_1 + q\mathbf{a}_2) - \mathrm{i}\omega t)$, we define a polarization parameter

$$\psi_{\mathbf{k}} = (\psi_{\mathrm{R}}, \psi_{\mathrm{L}}, \psi_{\mathrm{T}}) = \frac{1}{3R^2\left(\hat{\phi}_0\right)^2 + |\hat{\mathbf{u}}_0|^2}\left(3R^2\left(\hat{\phi}_0\right)^2, (\hat{\mathbf{u}}_0 \cdot \mathbf{e}_{\mathbf{k}})^2, (\hat{\mathbf{u}}_0 \cdot \mathbf{e}'_{\mathbf{k}})^2\right). \quad (13)$$

In which, $\mathbf{e}_{\mathbf{k}}$ ($\mathbf{e}'_{\mathbf{k}}$) represents a unit vector parallel (orthogonal) to the wave vector $\mathbf{k}$. The first component of $\psi_{\mathbf{k}}$ represents the participation ratio of the rotational mode, while the second and third components correspond to the participation ratios of the longitudinal mode and transverse mode, respectively. This polarization parameter is treated as a color represented by three RGB values. In this color scheme, red, green, and blue correspond to purely rotational, longitudinal, and transverse modes, respectively, whereas intermediate colors indicate mode hybridization.

The presence of zero modes has a significant impact on the wave characteristics of the metamaterial. A notable feature is that the metamaterial exhibits three rather than two acoustic branches emerging from the $\boldsymbol{\Gamma}$ point. This behavior arises directly from the pure rotation zero mode (Fig. 2(b)(i)), which can be interpreted as a Bloch mode with zero wavenumber. These three acoustic branches are accurately captured by the micropolar-continuum theory (see the three solid lines), with quantitative agreement observed over a moderate range of wave vectors. In contrast, the classical Cauchy-continuum theory predicts only two acoustic branches (shown as dashed lines), and its agreement with the discrete metamaterial calculation is only partial.

Another intriguing feature is the zero-frequency flat band along the $\boldsymbol{\Gamma}$-$\mathbf{M}$ direction, which originates from the other two zero modes of the metamaterial (Fig. 2(b)(ii) and (iii)). This zero-frequency band has previously been attributed to the so-called Guest–Hutchinson mode [23]. Notably, the flat band can also be effectively described with the micropolar-continuum theory. We assume a Bloch mode of the form $(\hat{\mathbf{u}}_0, \hat{\phi}_0)\exp(\mathrm{i}\mathbf{k}\cdot\mathbf{r})$, with $\hat{\mathbf{u}}_0$ and $\hat{\phi}_0$ being constant displacement and rotation. The micropolar strain and curvature can be derived as

$$\left(\frac{\mathrm{i}k_x\hat{u}_0 + \mathrm{i}k_y\hat{v}_0}{2}, \frac{\mathrm{i}k_x\hat{u}_0 - \mathrm{i}k_y\hat{v}_0}{2}, \frac{\mathrm{i}k_x\hat{v}_0 + \mathrm{i}k_y\hat{u}_0}{2}, \frac{\mathrm{i}k_x\hat{v}_0 - \mathrm{i}k_y\hat{u}_0}{2} - \hat{\phi}_0, \mathrm{i}k_x\hat{\phi}_0, \mathrm{i}k_y\hat{\phi}_0\right)\exp(\mathrm{i}\mathbf{k}\cdot\mathbf{r})\,. \quad (14)$$

It can be verified that along the $\boldsymbol{\Gamma}$-$\mathbf{M}$ direction a Bloch wave $(\hat{\mathbf{u}}_0, \hat{\phi}_0)\exp(\mathrm{i}\mathbf{k}\cdot\mathbf{r}) = (-1/2, \sqrt{3}/2, -1/R)\exp(\mathrm{i}\mathbf{k}\cdot\mathbf{r})$ induces the following micropolar strain and curvature $(\varepsilon_0, \varepsilon_\mathrm{d}, \varepsilon_\mathrm{s}, \varepsilon_\mathrm{a}, \kappa_1, \kappa_2) = (0, -\mathrm{i}\sqrt{3}/4, \mathrm{i}/4, \mathrm{i}/2 + 1/R, -\mathrm{i}\sqrt{3}/(2R), -\mathrm{i}/(2R))\exp(\mathrm{i}\mathbf{k}\cdot\mathbf{r})$, which is exactly a linear combination of the three zero modes shown in Fig. 2(b). Consequently, the Bloch wave frequency must be zero.

The polarization of waves in this metamaterial is complicated by the hybridization between rotational and translational DOFs. Figure 3(b) shows three modes at a small wave vector $\mathbf{k} = 0.1\mathbf{M}$ ($0.1\mathbf{K}$) along the $\boldsymbol{\Gamma}$-$\mathbf{M}$ ($\boldsymbol{\Gamma}$-$\mathbf{K}$) direction. The allowed hybridization between modes can be analyzed using group theory analysis. For wave propagation along the $\boldsymbol{\Gamma}$-$\mathbf{M}$ direction, the transverse and rotational modes are odd under mirror reflection with mirror lines along the $\boldsymbol{\Gamma}$-$\mathbf{M}$ direction, whereas the longitudinal mode is even under this symmetry operation. Modes with different parity cannot hybridize. Consequently, the first (Fig. 3(b), A1) and third (Fig. 3(b), A3) modes exhibit hybridization between rotation and transverse displacement, while the longitudinal mode remains a pure mode (see the green dots along the $\boldsymbol{\Gamma}$-$\mathbf{M}$ direction). The polarizations of the three acoustic branches of the metamaterial are also accurately captured by the micropolar-continuum theory (compare the colors of the dots and solid lines in Fig. 3(a)).

Results for the twisted Kagome lattice (Fig. 2(c)) are shown in Figs. 3(c) and 3(d). As previously demonstrated [5, 8], a nonzero twist angle $\theta \neq 0$ eliminates the zero-frequency band along the $\boldsymbol{\Gamma}$-$\mathbf{M}$ direction. The branch corresponding to rotational modes (red line in Fig. 3(c); Fig. 3(d), A3 and B3) is also lifted to a finite frequency. Owing to the large frequency separation between the rotational branch and the two acoustic branches, hybridization between rotational and acoustic modes becomes relatively weak. Due to the Poisson's ratio of $-1$, the Cauchy-continuum theory predicts two degenerated longitudinal and transverse branches. This trend agrees well with the metamaterial results in the long-wavelength limit (compare dashed lines with dots). At slightly larger wavenumbers, the two acoustic branches, particularly along the $\mathbf{\Gamma}$-$\mathbf{M}$ direction, exhibit obvious frequency splitting. This frequency splitting results from hybridization between the longitudinal and transverse modes, due to the broken mirror-reflection symmetry or chirality of the metamaterial. Our micropolar-continuum model accurately reproduces the splitting (compare solid lines with dots along the $\mathbf{\Gamma}$-$\mathbf{M}$ direction in Fig. 3(c)), which scales quadratically to wavenumber $\Delta\omega \approx \left|\sqrt{\hat{C}_{11}}\hat{B}_{21} + \sqrt{\hat{C}_{44}}\hat{B}_{22}\right| |\mathbf{k}|^2 / \left(2\sqrt{\rho\hat{C}_{22}\hat{C}_{44}}\right)$ following micropolar theory. Chirality leads to two elliptically polarized acoustic modes, exhibiting counterclockwise (Fig. 3(d), A1) and clockwise translation (Fig. 3(d), A2), respectively, which might find applications in mode conversion for longitudinal and transverse waves. The chiral modes are analogous to the circularly polarized modes observed in three-dimensional chiral metamaterials, which can also be described using micropolar theory [77].

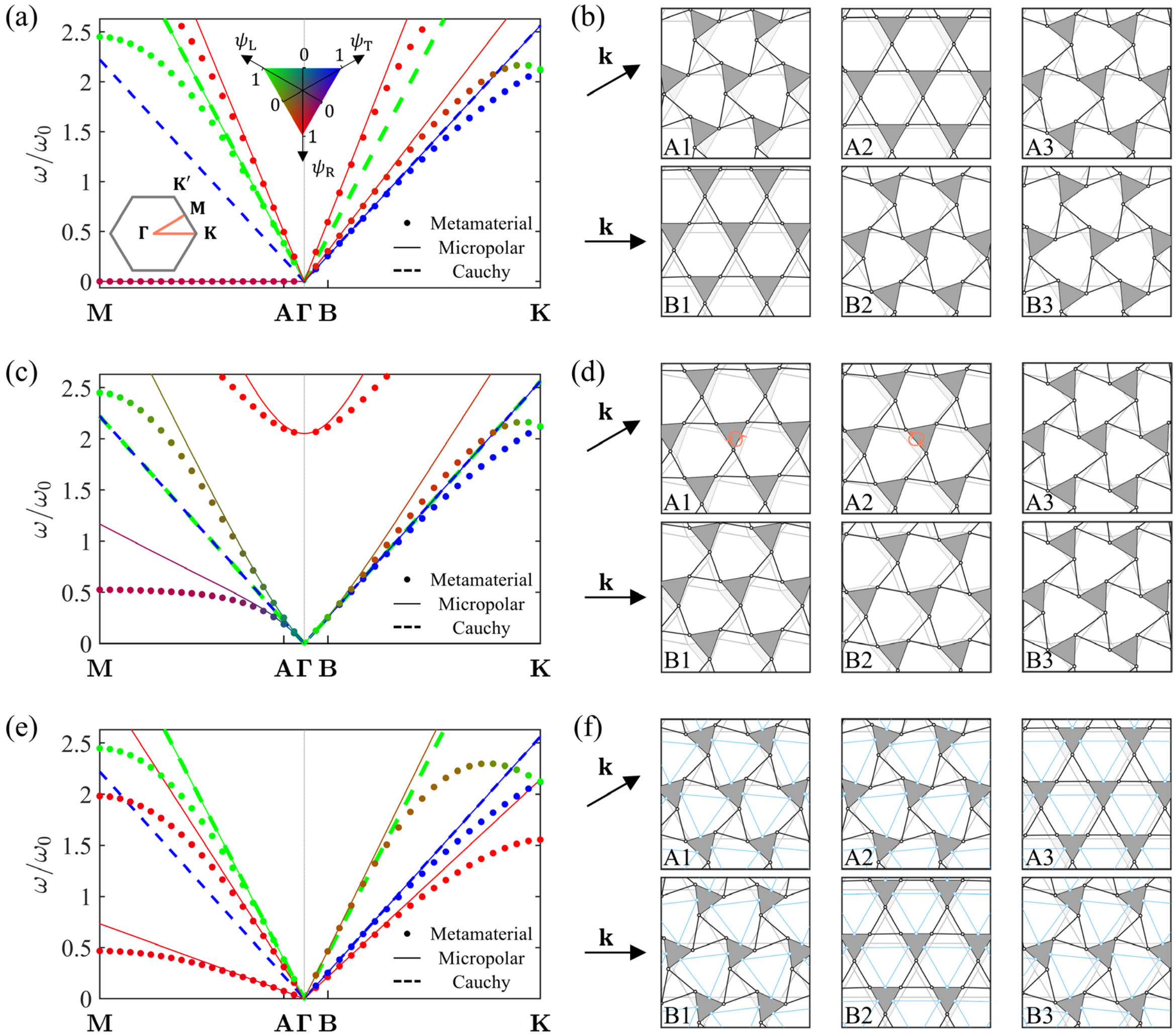

**Figure 3. Band structure and eigenmodes of the designed metamaterial shown in Fig. 2.** (a) Dispersion relations along high-symmetry direction (see inset). Results for the metamaterials are shown by symbols, while those for the micropolar-continuum (Cauchy-continuum) modeling are shown by solid (dashed) lines. Colors indicate the participation ratios of the transverse, longitudinal, and rotational components in each eigenmode, as defined in Eq. (13). Pure green, blue, and red

correspond to purely longitudinal, transverse, and rotational modes, respectively, whereas intermediate colors stand for hybridized modes. (b) Top three panels show the three eigenmodes for a small wave vector $\mathbf{k} = 0.1\mathbf{M}$ along the $\mathbf{\Gamma}$-$\mathbf{M}$ direction, as marked by letter **A** in (a). Bottom three panels correspond to a small wave vector $\mathbf{k} = 0.1\mathbf{K}$ along the $\mathbf{\Gamma}$-$\mathbf{K}$ direction, as marked by **B** in (a). (c) – (f) Same as (a) – (b) but for the metamaterials shown in Figs. 2(c) and (d), respectively.

We show in Figs. 3(e) and (f) the results for the metamaterial shown in Fig. 2(d), which possesses only one pure rotational zero mode. As discussed above, the rotational zero mode inevitably gives rise to three acoustic branches emerging from the $\mathbf{\Gamma}$ point. For wave propagation along the $\mathbf{\Gamma}$-$\mathbf{M}$ direction, the transverse and rotational modes are strongly coupled (see the magenta lines and magenta dots in Fig. 3(e)). Both modes display pronounced rotations (Fig. 3(f), A1 and A2) and transverse motion. Interestingly, if only the displacement DOFs are accessible from outside, the metamaterial appears to support two transverse modes rather than one transverse mode like usual 2D elastic metamaterials. Even in the long-wavelength limit, this hybridization persists due to the ideal zero mode.

According to Cauchy elasticity, the $\mathrm{C}_{3\mathrm{v}}/\mathrm{C}_3$ symmetric metamaterials discussed above are expected to be isotropic in the long-wavelength limit [72]. However, the dispersion relations shown in Fig. 3 shows strongly anisotropy arising from rotation-related zero modes. In the following, we further study wave propagations in these metamaterials and compare results obtained from Cauchy-continuum and micropolar-continuum models. As mentioned above, we focus on the metamaterial shown in Fig. 2(d) due to mechanical stability reasons.

Figure 4(a) presents the iso-frequency curves of the three acoustic branches for the metamaterial at $\omega/\omega_0 = 0.2$, which is chosen as the excitation frequency in the following wave simulations. Despite the metamaterial only exhibits three-fold rotation symmetry, the iso-frequency contours exhibit sixfold rotational symmetry because of the time-reversal symmetry, which ensures $\omega_i(\mathbf{k}) = \omega_i(-\mathbf{k})$ for the $i$-th band. The anisotropic iso-frequency curves are quantitatively captured by the micropolar-continuum model (Fig. 4(b)). Mathematically, the anisotropy results from a non-zero coupling tensor $B_{\alpha\beta\gamma}$ for the $\mathrm{C}_{3\mathrm{v}}$-symmetric metamaterial (see Table 1 and the effective micropolar elasticity matrix Eq. (C6) in **Appendix C**). In contrast, the Cauchy-continuum model predicts only two isotropic iso-frequency curves (Fig. 4(c)).

We first consider harmonic longitudinal wave excitation, implemented by imposing a harmonic radial displacement on a circular inner boundary (see the insets in Figs. 4(d)–4(f)). In this case, only the longitudinal-dominated mode is efficiently excited. We therefore plot the radial displacement component $u_\mathrm{r}$. As shown in Fig. 4(d), wave propagation in the metamaterial is highly localized along six directions, aligned with the high-symmetry $\mathbf{\Gamma}$ - $\mathbf{K}$ and $\mathbf{\Gamma}$ - $\mathbf{K}'$ directions. These preferred propagation directions can be inferred from the group velocity of the longitudinal-dominated mode, i.e., the normal direction of the corresponding iso-frequency contour (see the green curve in Fig. 4(a)). This pronounced anisotropic wave behavior is reproduced remarkably well by the micropolar-continuum model, whereas the Cauchy-continuum predicts a qualitatively different isotropic response. In all simulations, perfectly matched layers (PMLs) are employed to suppress reflections from the boundaries, mimicking an infinite large simulation domain. For the micropolar-continuum modeling, a mathematical PML based on the asymmetric transformation method is adopted. We remark that the global stiffness matrix of the metamaterial is singular due to the presence of rotational zero modes. However, in wave simulations, the global mass matrix is also involved, which stabilizes the simulations.

For transverse wave excitation, we again observe excellent agreement between the metamaterial simulation (Fig. 4(g)) and the micropolar-continuum prediction (Fig. 4(h)). In this case, transverse modes propagating along both the $\mathbf{\Gamma}$-$\mathbf{M}$ and $\mathbf{\Gamma}$-$\mathbf{K}$ directions are excited, leading to a pronounced interference pattern (see Figs. 4(g)(h)).

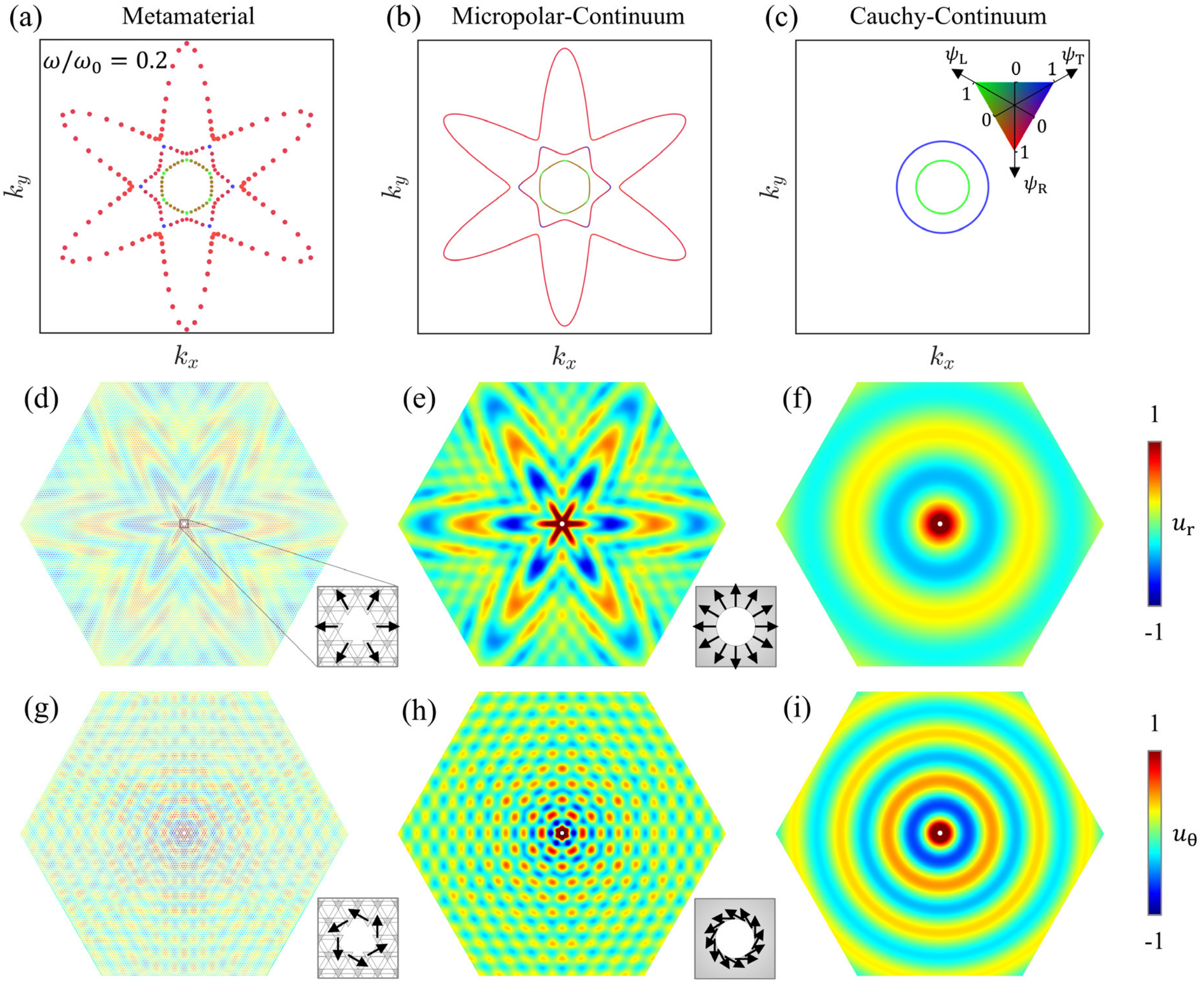

**Figure 4. Anisotropic wave propagation in the designed zero-mode micropolar metamaterials**. (a) Iso-frequency contours of the metamaterial in Fig. 2(d). Colors indicate the participation ratios of the transverse, longitudinal, and rotational components in each eigenmode. (b) and (c) results for the micropolar-continuum and Cauchy-continuum modeling. (d) Displacement field obtained with harmonic radial displacement excitation applied along a circular inner boundary (see inset), at the frequency $\omega/\omega_0 = 0.2$. False color represents the radial displacement component $u_r$. (e) and (f) results for the micropolar-continuum and Cauchy-continuum models, respectively. (g) – (i) Same as (d) – (f) but for transverse wave excitation, imposed by circumferential displacement along a circular boundary (see insets). The azimuthal component $u_\theta$ is plotted.

The pronounced anisotropic behavior of the metamaterial persists over a broad frequency range. As illustrative examples, we present results for longitudinal wave excitation at a lower frequency $\omega/\omega_0 = 0.1$ and a higher frequency $\omega/\omega_0 = 0.3$. In both cases, excellent agreement is observed between the discrete metamaterial simulations (Figs. 5(a) and 5(d)) and the micropolar-continuum predictions (Figs. 5(b) and 5(e)). We note that the unusual wave properties not only exist for discrete metamaterials with ideal zero-energy modes but also for realistic metamaterials with approximate zero modes. We have designed a solid metamaterial (Fig. D1(a) in **Appendix D**) following the discrete design in Fig. 2(d) and numerically shown that that the unusual behaviors preserve (see details in **Appendix D**).

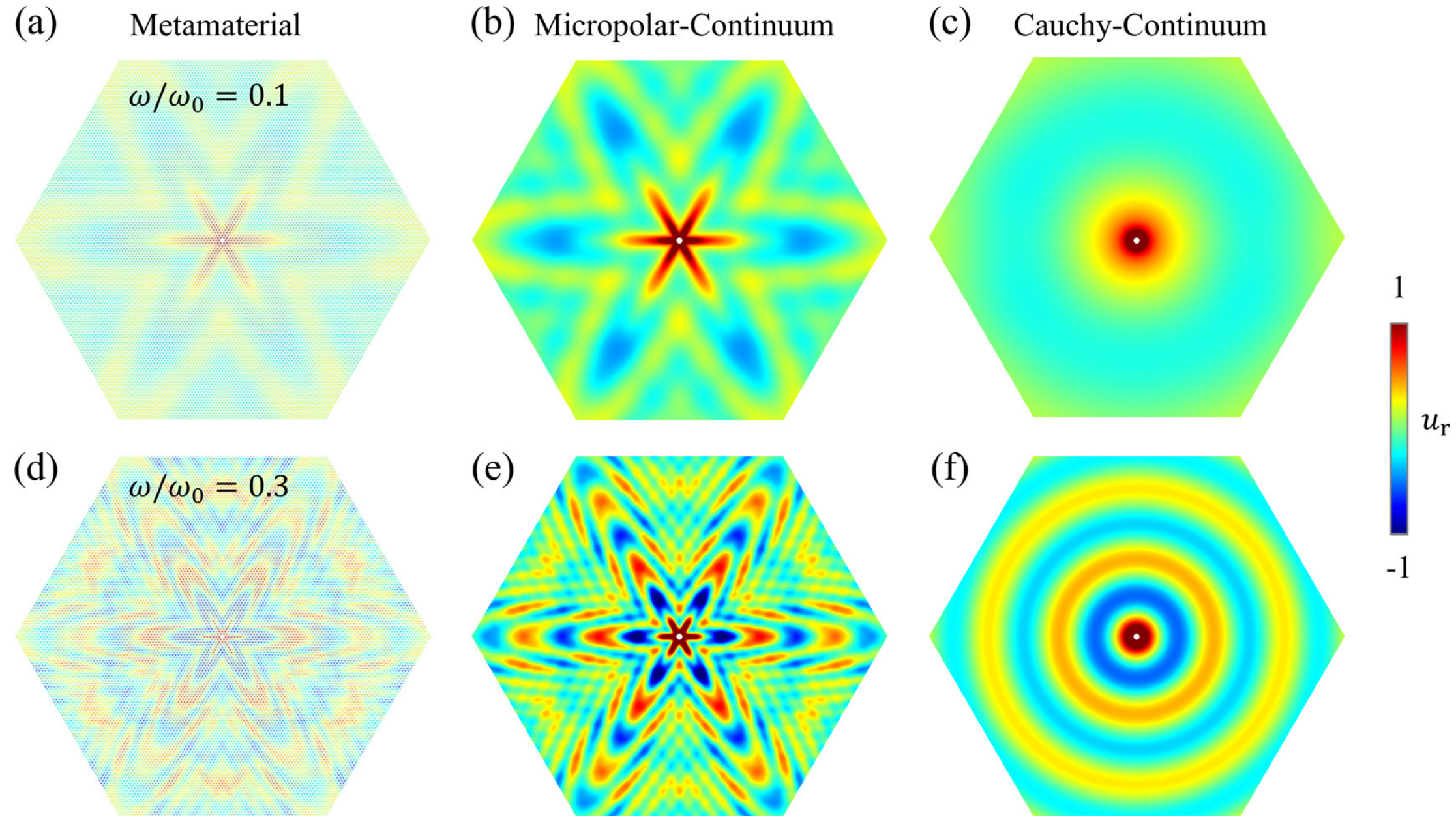

**Figure 5. Same as Fig. 4 with longitudinal excitation but at different excitation frequencies.** (a) – (c) $\omega/\omega_0 = 0.1$. (d) – (f) $\omega/\omega_0 = 0.3$.

The three acoustic modes of our designed metamaterial enable triple refraction beyond what is possible in the classical Cauchy media and conventional metamaterials. We consider an interface between a Cauchy-continuum domain on the left and a metamaterial domain on the right (see Fig. 6). We first investigate Gaussian longitudinal wave excitation at the same frequency $\omega/\omega_0 = 0.2$ as in Fig. 4. The incidence angle and the Cauchy-continuum parameters are chosen such that the three refraction angles are well separated. By applying Snell's law, i.e., conservation of the wavevector component parallel to the interface, the refraction angles of the three acoustic modes can be determined from the iso-frequency contours (Fig. 6(a)). In the numerically calculated displacement field (Fig. 6(b)), three refracted beams are clearly visible, although two of them are comparatively weaker. Their propagation directions agree precisely with those predicted by Snell's law (indicated by the cyan arrows). The insets show the displacement fields at selected locations along the three refracted beams, where the black arrows denote the wavevectors and the green arrows indicate the displacement polarizations. The strongest refracted beam corresponds to a longitudinal mode, whereas the negatively refracted beam results from a transverse-dominated mode. As before, the micropolar-continuum model (Fig. 6(c)) reproduces triple-refraction behavior, in excellent agreement with the metamaterial calculations.

For transverse wave excitation (Fig. 6(e)), the refraction angles can be analyzed similarly. In this case, the negatively refracted beam remains transverse-dominated and is more pronounced than the other two beams. Once again, the micropolar-continuum model accurately captures the triple-refraction phenomenon (Fig. 6(f)). We emphasize that unlike negative refraction associated with negative-index materials [84], both the negative refraction and the triple refraction here are broadband, because all of the three acoustic modes originate from the $\boldsymbol{\Gamma}$ point.

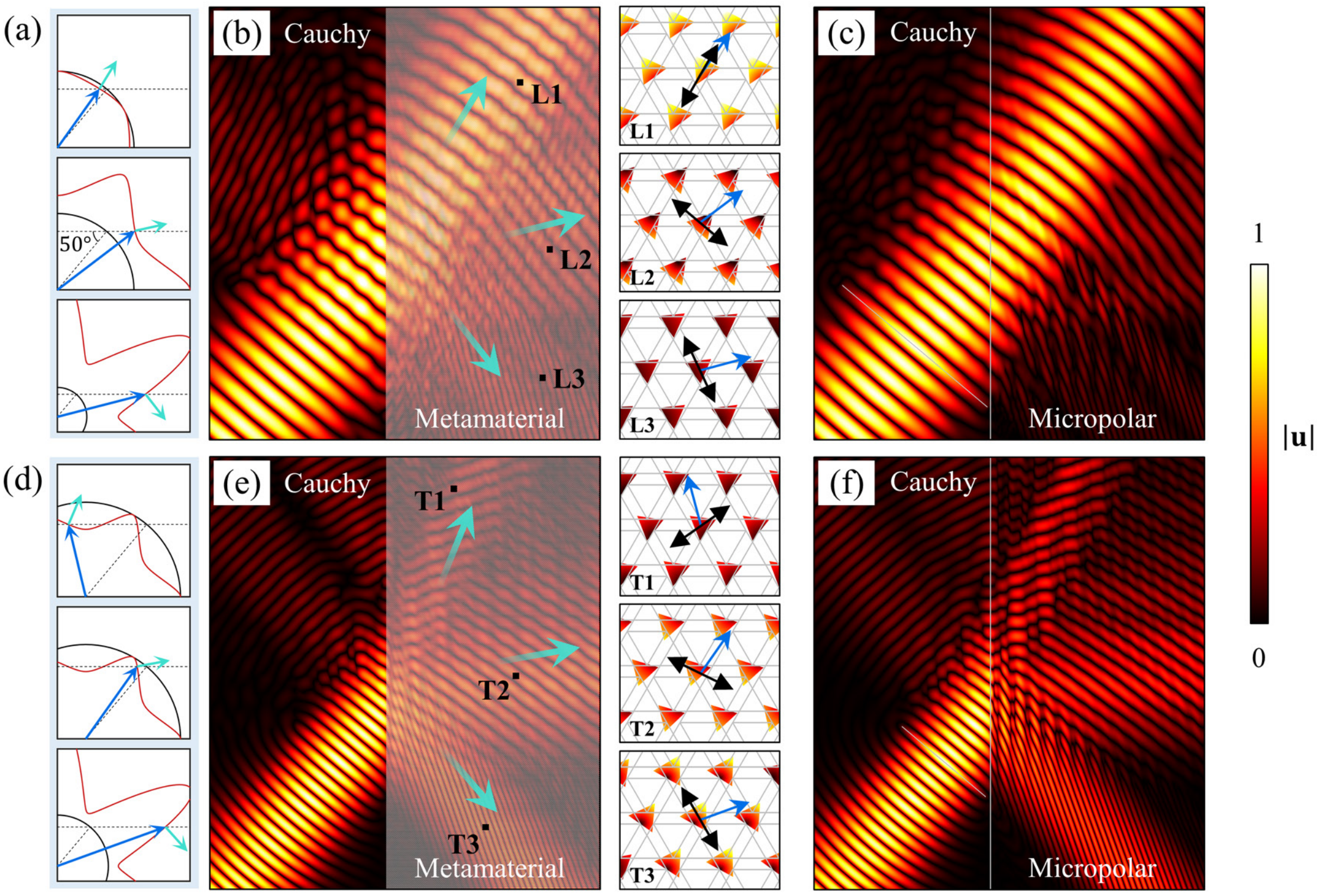

**Figure 6. Triple refraction in micropolar metamaterials with zero modes.** (a) Iso-frequency contours (red) of the metamaterial shown in Fig. 2(d) at the excitation frequency $\omega/\omega_0 = 0.2$. The three panels correspond to the three acoustic branches of the metamaterial. Black curves indicate the iso-frequency contours for the longitudinal branch of a background Cauchy medium. (b) Simulated displacement field for a longitudinal Gaussian beam incident from the background Cauchy medium (left) onto the metamaterial (right). The right three panels plot displacement fields at selected locations marked by the black squares. Blue and black arrows denote the wavevectors and displacement vectors, respectively. We choose the material parameters $\lambda = \mu = \sqrt{3}(s_1 + s_2)$, and $\rho = 2M/(\sqrt{3}a^2)$ for the Cauchy medium to clearly illustrate the triple refraction. (c) Same as (b) but using a micropolar-continuum model. (d) – (f) Same as (a) – (c) but for transverse wave incidence.

## Conclusions

In this paper, we have systematically classified zero modes, particularly those associated with internal rotations, within the 2D micropolar framework according to material symmetries. Based on this classification, we have successfully designed a metamaterial that supports several representative rotational micropolar zero modes, such as a pure rotation zero mode or a coupled zero mode from rotation and hydrostatic expansion. These rotational zero modes give rise to rich wave phenomena beyond the predictions of the classical Cauchy-continuum elasticity, including the emergence of three acoustic branches in the low-frequency limit, triple refraction, and chiral acoustic modes. We have verified the accuracy of micropolar-continuum theory in capturing these unconventional wave behaviors via band-structure calculations and simulations of wave propagation. This work provides guidance for exploration and modeling of other intriguing wave phenomena with rotational zero modes and may stimulate further experimental investigations of triple refraction.

## Appendix

**Appendix A**: Transformed Micropolar Elasticity Matrix

According to the strain bases defined in Eq. (6) and the stress bases defined in Eq. (8), the constitutive law in Eq. (4) can be re-formulated as in Eq. (10). The elements in the transformed elasticity matrix are related to the micropolar elastic tensors as

$$\begin{cases} \hat{C}_{11} = C_{1111} + 2C_{1122} + C_{2222}, \quad \hat{C}_{22} = C_{1111} - 2C_{1122} + C_{2222}, \quad \hat{C}_{12} = C_{1111} - C_{2222}, \\ \hat{C}_{33} = C_{1212} + 2C_{1221} + C_{2121}, \quad \hat{C}_{44} = C_{1212} - 2C_{1221} + C_{2121}, \quad \hat{C}_{34} = C_{1212} - C_{2121}, \\ \hat{C}_{13} = C_{1112} + C_{1121} + C_{2212} + C_{2221}, \quad \hat{C}_{14} = C_{1112} - C_{1121} + C_{2212} - C_{2221}, \\ \hat{C}_{23} = C_{1112} + C_{1121} - C_{2212} - C_{2221}, \quad \hat{C}_{24} = C_{1112} - C_{1121} - C_{2212} + C_{2221}, \\ \hat{B}_{11} = B_{111} + B_{221}, \quad \hat{B}_{21} = B_{111} - B_{221}, \quad \hat{B}_{31} = B_{121} + B_{211}, \quad \hat{B}_{41} = B_{121} - B_{211}, \\ \hat{B}_{12} = B_{112} + B_{222}, \quad \hat{B}_{22} = B_{112} - B_{222}, \quad \hat{B}_{32} = B_{122} + B_{212}, \quad \hat{B}_{42} = B_{122} - B_{212}, \\ \hat{A}_{11} = A_{11}, \quad \hat{A}_{12} = A_{12}, \quad \hat{A}_{22} = A_{22}. \end{cases} \tag{A1}$$

**Appendix B**: Derivation of Zero Modes of the Designed Metamaterials

For the metamaterial shown in Fig. 2(b), the strain energy associated to the unit cell $(p, q)$ consists of potential energy in the six black springs. This energy can be expressed as

$$U_{p,q} = \frac{1}{2} V_{p,q} = \frac{1}{4} s_1 \sum_{n=1}^{6} e_n^2 \,. \tag{B1}$$

The pre-factor $1/2$ appears because the energy stored in each spring is shared by its two neighboring cells. $e_n$ is the elongation of the $n$-th spring connected to the central mass block (see Fig. B1). Elongation of each spring can be derived from the displacements of both ends. For example, $e_1$ can be written as

$$e_1 = (\mathbf{u}_{p+1,q} + \phi_{p+1,q} R \mathbf{r}_2' - \mathbf{u}_{p,q} - \phi_{p,q} R \mathbf{r}_1') \cdot \mathbf{d}_1. \tag{B2}$$

Similarly, the elongations of other springs can be obtained. $\mathbf{d}_i$ and $\mathbf{r}_i$ are unit vectors indicated in Fig. B1, and $\mathbf{r}_i'$ is obtained from $\mathbf{r}_i$ by a rotation of 90°.

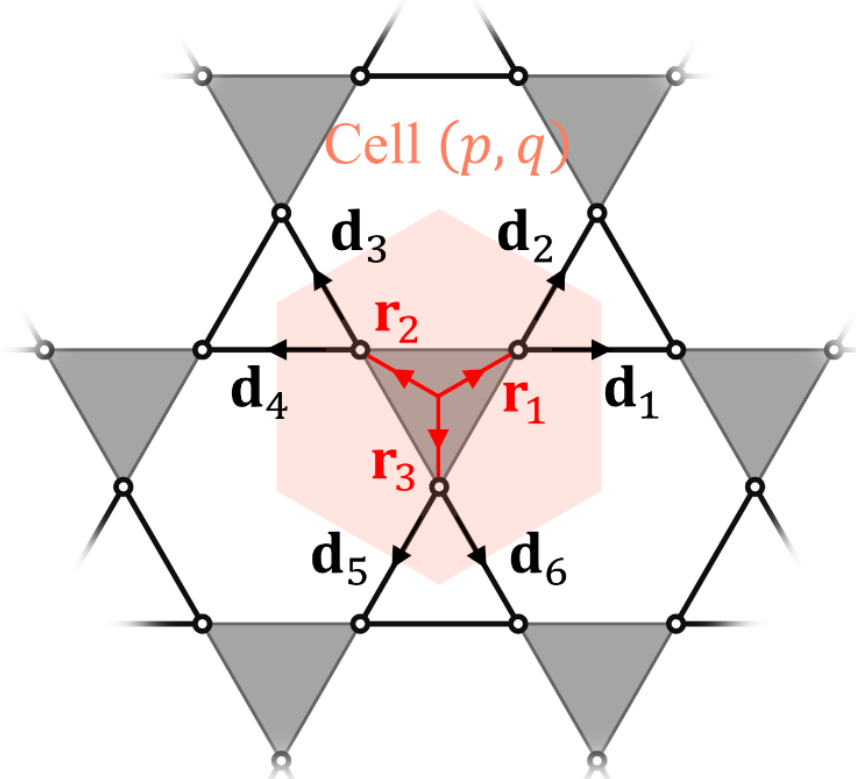

**Figure B1**. **Vectors used for calculation strain energy for the metamaterial in Fig. 2(b).** We denote the unit vectors along the six springs connected to the central mass block by $\mathbf{d}_i$, $i = 1, 2 \dots 6$. $\mathbf{r}_i$ with $i$ ranging from 1 to 3 stand for unit vectors pointing from the mass block's center to its three corners.

We can check under what deformations the strain energy is zero. Here, we are interested in affine deformations as they can be described by the effective-medium theory. These affine deformations are the long-wavelength limit solutions. We assume the following displacement and rotation

$$\begin{cases} \mathbf{u}_{p+m,q+n} = \mathbf{u}_{p,q} + \mathbf{u}\nabla \cdot (m\mathbf{a}_1 + n\mathbf{a}_2) \\ \phi_{p+m,q+n} = \phi_{p,q} + \phi\nabla \cdot (m\mathbf{a}_1 + n\mathbf{a}_2) \end{cases}. \tag{B3}$$

In which, $\mathbf{u}\nabla$ should be understood as a constant second-order tensor and $\phi\nabla$ is a constant vector. The affine deformation is uniquely determined by the constant second-order tensor and vector. Elongations of the six springs (see Fig. B1) induced by the deformation in Eq. (B3) are

$$\begin{pmatrix} e_1 \\ e_2 \\ e_3 \\ e_4 \\ e_5 \\ e_6 \end{pmatrix} = \mathbf{D} \begin{pmatrix} \mathbf{u}\nabla \cdot \mathbf{a}_1 \\ \mathbf{u}\nabla \cdot \mathbf{a}_2 \\ \phi_{p,q} \\ \phi\nabla \cdot \mathbf{a}_1 \\ \phi\nabla \cdot \mathbf{a}_2 \end{pmatrix} = \begin{pmatrix} \mathbf{d}_1 & \mathbf{0} & \mathbf{d}_1 \cdot R(\mathbf{r}_2' - \mathbf{r}_1') & \mathbf{d}_1 \cdot R\mathbf{r}_2' & 0 \\ \mathbf{0} & \mathbf{d}_2 & \mathbf{d}_2 \cdot R(\mathbf{r}_3' - \mathbf{r}_1') & 0 & \mathbf{d}_2 \cdot R\mathbf{r}_3' \\ -\mathbf{d}_3 & \mathbf{d}_3 & \mathbf{d}_3 \cdot R(\mathbf{r}_3' - \mathbf{r}_2') & -\mathbf{d}_3 \cdot R\mathbf{r}_3' & \mathbf{d}_3 \cdot R\mathbf{r}_3' \\ -\mathbf{d}_4 & \mathbf{0} & \mathbf{d}_4 \cdot R(\mathbf{r}_1' - \mathbf{r}_2') & -\mathbf{d}_4 \cdot R\mathbf{r}_1' & 0 \\ \mathbf{0} & -\mathbf{d}_5 & \mathbf{d}_5 \cdot R(\mathbf{r}_1' - \mathbf{r}_3') & 0 & -\mathbf{d}_5 \cdot R\mathbf{r}_1' \\ \mathbf{d}_6 & -\mathbf{d}_6 & \mathbf{d}_6 \cdot R(\mathbf{r}_2' - \mathbf{r}_3') & \mathbf{d}_6 \cdot R\mathbf{r}_2' & -\mathbf{d}_6 \cdot R\mathbf{r}_2' \end{pmatrix} \begin{pmatrix} \mathbf{u}\nabla \cdot \mathbf{a}_1 \\ \mathbf{u}\nabla \cdot \mathbf{a}_2 \\ \phi_{p,q} \\ \phi\nabla \cdot \mathbf{a}_1 \\ \phi\nabla \cdot \mathbf{a}_2 \end{pmatrix}. \tag{B4}$$

The compatibility matrix $\mathbf{D}$ is $6 \times 7$. Substituting $\mathbf{r}_i'$ and $\mathbf{d}_i$ into the matrix $\mathbf{D}$, it can be verified that the matrix $\mathbf{D}$ has a 4-dimensional null space

$$\mathbf{N} = \begin{pmatrix} 0 & 0 & 1 & 0 \\ 1 & 1 & \sqrt{3} & 0 \\ -\frac{\sqrt{3}}{2} & -\frac{\sqrt{3}}{2} & 0 & \sqrt{3} \\ \frac{1}{2} & \frac{1}{2} & 0 & 0 \\ 1 & -1 & 0 & 0 \\ 0 & 0 & \frac{2}{R} & 0 \\ 0 & 0 & 0 & -\frac{\sqrt{3}}{R} \end{pmatrix}, \mathbf{DN} = \mathbf{0}. \tag{B5}$$

The first column of $\mathbf{N}$ corresponds to a rigid-body rotation and represents a trivial zero mode. The remaining three columns stand for non-trivial zero modes. When projected onto the deformation bases defined in Eq. (6)(7), the zero modes are $(\varepsilon_{\mathrm{o}}, \varepsilon_{\mathrm{d}}, \varepsilon_{\mathrm{s}}, \varepsilon_{\mathrm{a}}, \kappa_1, \kappa_2) = (0, 0, 0, 1, 0, 0)$, $(0, 1, 0, 0, 2/R, 0)$, and $(0, 0, 1, 0, 0, -2/R)$, respectively.

**Appendix C**: Derivation of Effective Micropolar-Continuum and Cauchy-Continuum Parameters for the Designed Metamaterials

For the metamaterial shown in Fig. 2(a), the strain energy density can be obtained from the strain energy in the representative unit cell $(p, q)$

$$w = \frac{1}{4S_{\mathrm{cell}}} \sum_{n=1}^{6} (s_1 e_n^2 + s_2 l_n^2) \,. \tag{C1}$$

The elongations of the black and blue springs are denoted by $e_n$ and $l_n$, respectively. We assume a Taylor expansion of the displacement and the rotation

$$\begin{cases} \mathbf{u}_{p+m,q+n} = \mathbf{u}_{p,q} + \mathbf{u}\nabla \cdot \mathrm{d}\mathbf{x} + \frac{1}{2}\mathrm{d}\mathbf{x} \cdot \nabla\mathbf{u}\nabla \cdot \mathrm{d}\mathbf{x} + \cdots \\ \phi_{p+m,q+n} = \phi_{p,q} + \phi\nabla \cdot \mathrm{d}\mathbf{x} + \frac{1}{2}\mathrm{d}\mathbf{x} \cdot \nabla\phi\nabla \cdot \mathrm{d}\mathbf{x} + \cdots, \end{cases} \tag{C2}$$

with $\mathrm{d}\mathbf{x} = m\mathbf{a}_1 + n\mathbf{a}_2$. The strain energy density can be obtained by substituting Eq. (C2) into Eq. (C1). A strain energy density function $w(\nabla\mathbf{u}, \phi, \nabla\phi\,)$ can be obtained by truncating the Taylor expansion to first order. By comparing the micropolar energy density (Eq. (3)), one can obtain a set of effective micropolar parameters for the metamaterial. Alternatively, one can obtain the effective parameters as

$$C_{\alpha\beta\gamma\delta} = \frac{\partial^2 w(\nabla\mathbf{u}, \phi, \nabla\phi)}{\partial u_{\beta,\alpha} \partial u_{\delta,\gamma}}, \;\; B_{\alpha\beta\gamma} = \frac{\partial^2 w(\nabla\mathbf{u}, \phi, \nabla\phi)}{\partial u_{\beta,\alpha} \partial \phi_{,\gamma}}, \;\; A_{\alpha\beta} = \frac{\partial^2 w(\nabla\mathbf{u}, \phi, \nabla\phi)}{\partial \phi_{,\alpha} \partial \phi_{,\beta}}. \tag{C3}$$

The transformed elasticity matrix can be obtained following Eq. (A1).

For the metamaterial shown in Fig. 2(b), the transformed effective micropolar elasticity matrix is

$$\begin{pmatrix} \hat{\mathbf{C}} & \hat{\mathbf{B}} \\ \hat{\mathbf{B}}^{\mathrm{T}} & \hat{\mathbf{A}} \end{pmatrix} = \frac{\sqrt{3}}{4} \begin{pmatrix} 8s_1 & 0 & 0 & 0 & 0 & 0 \\ & 4s_1 & 0 & 0 & -2s_1 R & 0 \\ & & 4s_1 & 0 & 0 & 2s_1 R \\ & & & 0 & 0 & 0 \\ & \mathrm{sym} & & & s_1 R^2 & 0 \\ & & & & & s_1 R^2 \end{pmatrix}, \tag{C4}$$

which has three zero eigenvalues, corresponding to three zero modes consistent with those derived in **Appendix B**.

The transformed effective elasticity matrix of the metamaterial shown in Fig. 2(c) is

$$\frac{\sqrt{3}}{4} \begin{pmatrix} 8s_1 \cos^2\theta & 0 & 0 & -8s_1 \sin\theta\cos\theta & 0 & 0 \\ & 4s_1 & 0 & 0 & -2s_1 R\cos\theta\cos 2\theta & 2s_1 R\sin\theta\cos 2\theta \\ & & 4s_1 & 0 & 2s_1 R\sin\theta\cos 2\theta & 2s_1 R\cos\theta\cos 2\theta \\ & & & 8s_1 \sin^2\theta & 0 & 0 \\ & \mathrm{sym} & & & s_1 R^2(2-\cos 4\theta) & 0 \\ & & & & & s_1 R^2(2-\cos 4\theta) \end{pmatrix}, \tag{C5}$$

which has one coupled zero mode, $(\varepsilon_{\mathrm{o}}, \varepsilon_{\mathrm{d}}, \varepsilon_{\mathrm{s}}, \varepsilon_{\mathrm{a}}, \kappa_1, \kappa_2) = (\sin\theta, 0, 0, \cos\theta, 0, 0)$.

For the metamaterial shown in Fig. 2(d), the transformed effective elasticity matrix is

$$\frac{\sqrt{3}}{4} \begin{pmatrix} 8(s_1+s_2) & 0 & 0 & 0 & 0 & 0 \\ & 4(s_1+s_2) & 0 & 0 & (s_2-2s_1)R & 0 \\ & & 4(s_1+s_2) & 0 & 0 & (2s_1-s_2)R \\ & & & 0 & 0 & 0 \\ & \mathrm{sym} & & & \left(s_1+\frac{s_2}{4}\right)R^2 & 0 \\ & & & & & \left(s_1+\frac{s_2}{4}\right)R^2 \end{pmatrix}, \tag{C6}$$

which has a pure rotation zero mode, $(\varepsilon_{\mathrm{o}}, \varepsilon_{\mathrm{d}}, \varepsilon_{\mathrm{s}}, \varepsilon_{\mathrm{a}}, \kappa_1, \kappa_2) = (0, 0, 0, 1, 0, 0)$.

In order to derive effective Cauchy-continuum parameters for the metamaterials in Fig. 2, we cannot treat the rotational DOF, $\phi_{p,q}$, as an independent macroscopic kinetic variable. Therefore, we assume $\phi_{p,q}$ is the same for all unit cells and solve the rotation by minimizing the potential energy $V_{p,q}$ with respect to $\phi_{p,q}$. In this way, $\phi_{p,q}$ can be represented by displacement DOFs.

Then, we consider the deformation corresponding to a given constant Cauchy strain tensor $\boldsymbol{\varepsilon}$. The displacement can be assumed as, following the Cauchy-Born condition

$$\mathbf{u}_{p+m,q+n} = \mathbf{u}_{p,q} + \boldsymbol{\varepsilon} \cdot (m\mathbf{a}_1 + n\mathbf{a}_2). \tag{C7}$$

Substitute the displacement field and the solved rotational DOFs into the strain energy density defined in Eq. (C1), one can obtain a strain energy density function $w$ in terms of the Cauchy strain tensor $\boldsymbol{\varepsilon}$. The Cauchy elasticity parameters are given by

$$C_{\alpha\beta\gamma\delta} = \frac{\partial^2 w(\boldsymbol{\varepsilon})}{\partial \varepsilon_{\alpha\beta} \partial \varepsilon_{\gamma\delta}} , \tag{C8}$$

The transformed Cauchy elasticity matrices of metamaterials in Fig. 2(b-d) are

$$\hat{\mathbf{C}}^{(\mathrm{b})} = \sqrt{3} s_1 \begin{pmatrix} 2 & & \\ & 1 & \\ & & 1 \end{pmatrix}, \ \hat{\mathbf{C}}^{(\mathrm{c})} = \sqrt{3} s_1 \begin{pmatrix} 0 & & \\ & 1 & \\ & & 1 \end{pmatrix}, \ \hat{\mathbf{C}}^{(\mathrm{d})} = \sqrt{3} (s_1 + s_2) \begin{pmatrix} 2 & & \\ & 1 & \\ & & 1 \end{pmatrix}. \tag{C9}$$

The strain bases and stress bases for Cauchy elasticity are defined the same as for micropolar elasticity, except the asymmetric strain and stress bases that are not needed. As mentioned in the main paper, the effective Cauchy-continuum elasticity matrices cannot reflect the zero modes of the metamaterials in Figs. 2(b) and (d).

**Appendix D**: Designed solid metamaterial and its wave behavior

For A solid metamaterial (Fig. D1(a)) is designed by replacing the springs in the discrete model in Fig. 2(d) with slender beams. The lattice vectors and the dimensions of the triangular mass blocks are kept the same as those of the discrete model. Detailed geometric parameters are provided in the **Data and Code Availability.**

Due to the finite contact areas between the beams and the triangular blocks, the pure rotational mode becomes a quasi-zero-energy mode rather than an ideal zero-energy mode. Numerical band structure (Fig. D1(b)) shows that the rotational branch has a finite cut-off frequency at zero wavenumber, due to the finite energy of the pure rotational mode. Nevertheless, we still obtain three dispersionless branches over a wide frequency range above the cut-off frequency. The wave speed along $\boldsymbol{\Gamma}\mathbf{M}$ and $\boldsymbol{\Gamma}\mathbf{K}$ are also rather different, in consistent with the discrete model in the main text. The frequency is normalized by $f_0 = c_{\mathrm{T}}/(20a)$, in which $a$ is the lattice constant and $c_{\mathrm{T}}$ is the long-wavelength limit speed for transverse wave along the $\boldsymbol{\Gamma}\mathbf{K}$ direction.

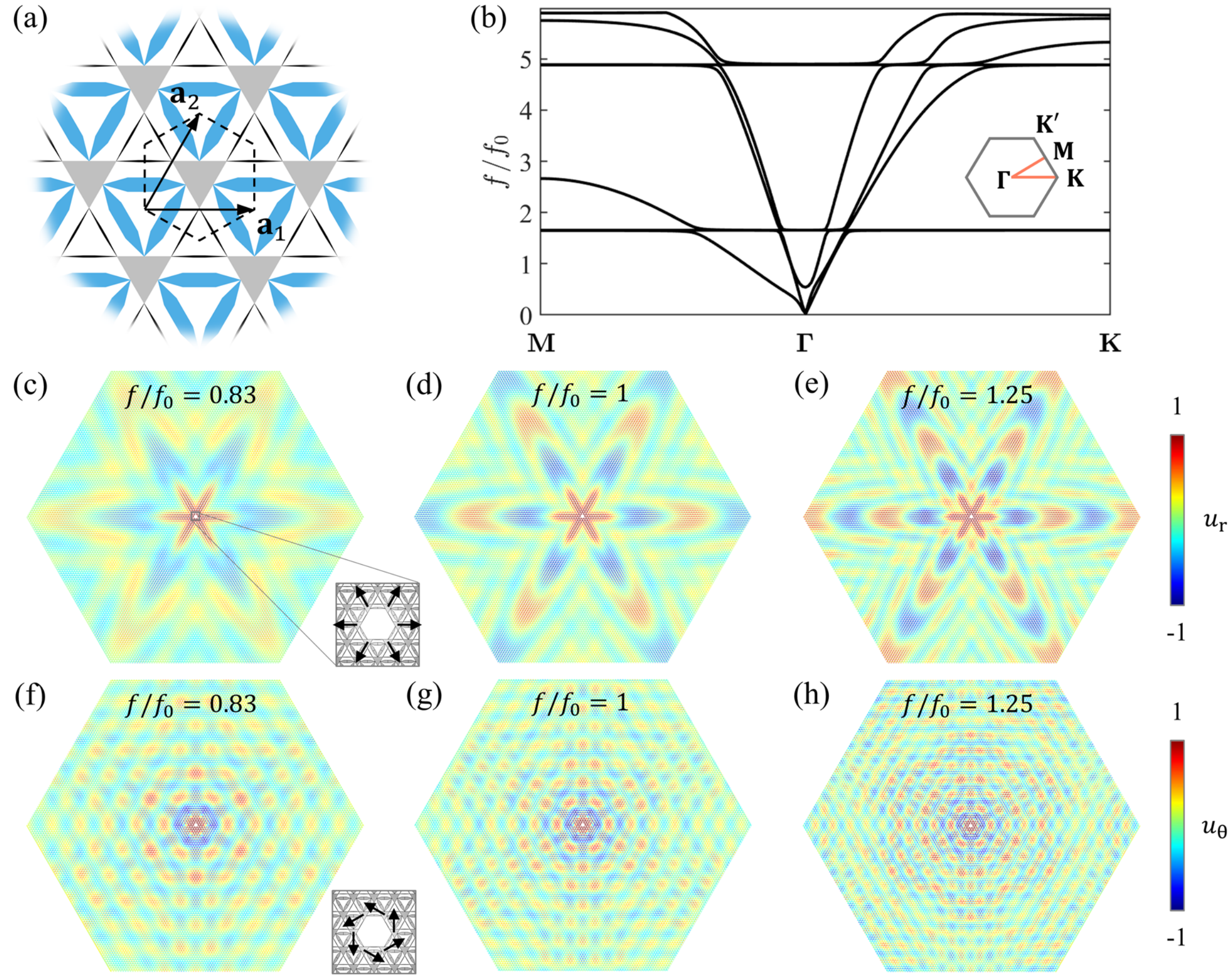

**Figure D1**. **Solid metamaterial model and its anisotropic wave behavior.** (a) Designed solid metamaterial model with a unit cell and two lattice vectors indicated. Material parameters are chosen as follows: $E_1 = 400\text{GPa}$, $\nu_1 = 0.29$, and $\rho_1 = 1.9 \times 10^4 \text{kg/m}^3$ for the gray triangular blocks, $E_2 = 205\text{GPa}$, $\nu_2 = 0.3$, and $\rho_2 = 7.85 \times 10^3 \text{kg/m}^3$ for the blue beams, and $E_3 = 20\text{GPa}$, $\nu_3 = 0.25$, and $\rho_3 = 1.6 \times 10^3 \text{kg/m}^3$ for the black slender beams. (b) Band structure of the metamaterial along high-symmetry directions (see inset). (c) Displacement field obtained with harmonic radial displacement (longitudinal) excitation applied along a circular inner boundary (see inset), at the frequency $f/f_0 = 0.83$. False color represents the radial displacement component $u_\text{r}$. (d) and (e) results for excitation at the frequency $f/f_0 = 1$ and $f/f_0 = 1.25$. (f) - (h) Same as (c) – (e) but for transverse wave excitation (see inset). The color represents the circumferential displacement component $u_\theta$.

We perform wave simulations as in Figs. 4 and 5 with longitudinal and transverse wave excitation for three different excitation frequencies, $f = 0.83f_0$, $f = 1.25f_0$, and $f = 1.00f_0$, respectively. The results are shown in Fig. D1(c) – (h). All these frequencies lie above the cut-off frequency of the rotational branch. These results confirm that the solid metamaterial exhibits anisotropic wave propagation over a broad low-frequency range, as our discrete model shows.

**Data and code availability**

The data that supports the plots within this paper are available on the open access data repository of Zenodo [https://doi.org/10.5281/zenodo.19878807].

**Acknowledgements**

This work is supported by the National Natural Science Foundation of China (Grant Nos. 12532006).

**Author contributions**

D.S. performed the theory analysis and numerical simulations. Y. C. drafted the paper. All authors contributed to the writing, review, and revision of the paper. Y.C. and G.K. supervised the project.

**Conflict of interests**

Authors declare that they have no competing interests.